# On the quest of low temperature nitrogen infusion relevant for superconducting Nb based radio-frequency cavities

G. D. L Semione[1,2*], A. D. Pandey [1], S. Tober[1,2], J. Pfrommer[1], A. Poulain[3], J. Drnec[3], G. Schütz[4], T. F. Keller[1,2], H. Noei[1], V. Vonk[1], B. Foster[1,2,5], A. Stierle[1,2]

[1]*Deutsches Elektronen-Synchrotron DESY, 22607 Hamburg, Germany*

[2]*Fachbereich Physik, Universität Hamburg, 22607 Hamburg, Germany*

[3]*European Synchrotron Radiation Facility, 71 avenue des Martyrs, CS 40220, Grenoble Cedex 9, France*

[4]*Max Planck Institute for Intelligent Systems, Heisenbergstrasse 1, D-70569 Stuttgart, Germany*

[5]*University of Oxford, Oxford, UK*

## ABSTRACT

A detailed study of the near-surface structure and composition of Nb, the material of choice for Superconducting Radio Frequency accelerator (SRF) cavities, is of great importance in order to understand the effects of different treatments applied during cavity production. By means of surface-sensitive techniques such as grazing incidence diffuse X-ray scattering, X-ray reflectivity and X-ray photoelectron spectroscopy, single-crystalline Nb(100) samples were investigated *in* and *ex-situ* during annealing in UHV as well as in nitrogen atmospheres with temperatures and pressures similar to the ones employed in real Nb cavity treatments. Annealing of Nb specimens up to 800°C in vacuum promotes partial reduction of the natural surface oxides ($Nb_2O_5$, $NbO_2$, $NbO$) into $NbO$. Upon cooling to 120°C, no evidence of nitrogen-rich layers was detected after nitrogen exposure times of up to 48 hours. Oxygen enrichment below the Nb/oxide interface and posterior diffusion of oxygen species towards the Nb matrix, along with a partial reduction of the natural surface oxides was observed upon a stepwise annealing up to 250°C. Nitrogen introduction to the system at 250°C neither promotes N diffusion into the Nb matrix nor the formation of new surface layers. Upon further heating to 500°C in a nitrogen atmosphere, the growth of a new subsurface $Nb_xN_y$ layer was detected. These results shed light on the composition of the near-surface region of Nb after low-temperature nitrogen treatments, which are reported to lead to a performance enhancement of SRF cavities.

# I. INTRODUCTION

Superconducting radio-frequency (SRF) cavities are the foundation of modern particle accelerators and free-electron lasers (*e. g* FLASH and EU-XFEL in Hamburg, Germany). Achieving higher accelerating gradients and better efficiency, as well as increasing their performance for continuous wave operation (CW) provides several challenges ranging from fundamental science up to large scale production[1–3]. Recently, thermal treatments performed at high (800°C-900°C) and low (120°C-160°C) temperatures and involving nitrogen presence were reported to improve the quality factor ($Q_0$ – the ratio of energy stored in a cavity to the energy lost per RF cycle) up to three times when compared with a regularly produced cavity[2,3]. However, an understanding of the processes involved and the parameters required is still elusive.

A type-II superconductor, niobium (Nb) has the highest critical temperature and critical field among all pure elements. Combined with its resistance to corrosion and ductility, Nb has become the material of choice for SRF-based technology. The concentration of impurities (hydrogen, oxygen, nitrogen and carbon) within the RF penetration layer (≈40 nm for pure Nb) is believed to be the key factor regarding the performance of cavities, either by lowering the electron mean-free-path[4], energy gap[5] or by trapping potential hydrogen which can cluster and precipitate niobium hydrides upon cooldown[6,7]. Traditionally, Nb-based SRF cavities were subjected to several production steps, including different polishing cycles like buffered chemical polishing (BCP) and electrochemical polishing (EP), as well as heat treatments at 800°C and 120°C[1,8] in high-vacuum in order to lower the hydrogen content inherent to the material[9]. In recent years, controlled annealing of SRF cavities in nitrogen atmospheres, either at high or low temperatures for different time spans, showed a decrease in their surface resistance[2,3]. The former treatment, known as nitrogen doping, includes annealing polycrystalline Nb cavities for a few minutes at 800°C in $3·10^{-2} – 8·10^{-2}$ mbar of $N_2$. Such treatment conditions lead to the formation of β-$Nb_2N$ inclusions at the surface as well as promoting the growth of NbO[6,10]. By etching approximately 5 μm by electrochemical polishing (EP), therefore removing the aforementioned inclusions, cavities displayed record-breaking quality factors at intermediate accelerating field gradients up to 20 MV/m. Another treatment, called nitrogen infusion, consists in annealing Nb cavities at 800°C for 3 hours in high-vacuum ($10^{-7} – 10^{-6}$ mbar) followed by a decrease in temperature to the range 120°C-200°C. Subsequently, nitrogen is injected at a pressure of $3·10^{-2}$ mbar. The system is then kept in such conditions for 48 hours, yielding cavities with quality factors in the range of $6·10^{10}$ and accelerating gradients up to 45 MVm$^{-1}$ without the need of material removal[2]. Low-temperature annealing of Nb cavities under $N_2$ atmosphere is therefore an intense topic of research, since operation at high accelerating gradients (>25 MVm$^{-1}$) with high $Q_0$ are necessary for potential upgrades of existing accelerators into CW operation, like the European X-ray free-electron laser (EU-XFEL), as well as the development of future particle accelerators projects. However, insights on the actual surface state during and after such treatments are lacking. Ideally, the knowledge obtained from model systems (*e. g* single-crystal surfaces) can potentially be applied to real cavity materials and eventually lead to the discovery of new cavity treatments that will improve the performance limits.

Most of the studies found in the literature address N and Nb interactions at high-temperatures[11–14]. Nonetheless, few reports detail investigations of Nb single-crystals under low $N_2$ pressures and of cavity-grade Nb after undergoing real cavity-treatment conditions. Exposure of an oxide-free Nb(100) surface to $N_2$ pressures in range $7·10^{-8} – 7·10^{-7}$ mbar at 620K lead to the formation of a NbN(100) epitaxial layer, while exposure at 300K formed NbN clusters at the surface[15]. For polycrystalline cavity-grade Nb, nanoscale niobium-nitride islands were reported after heat treatments between 120°C – 160°C with a $N_2$ pressure of ~$3 \cdot 10^{-2}$ mbar, while photoelectron measurements displayed the presence of $NbN_{(1-x)}O$ below the natural oxide layer. Secondary Ion Mass Spectrometry (SIMS) measurements revealed a diffusion of oxygen species for ~20 nm, while nitrogen diffused ~50 nm, having the highest concentration in the first few nm when the exposure temperature was 140°C[16].

In this study, we investigated *in* and *ex-situ* the effects of the recently proposed nitrogen infusion treatment in the near-surface region of Nb as well as the effect of stepwise annealing in UHV up to 500°C in nitrogen atmosphere by means of X-ray reflectivity (XRR), X-ray photoemission spectroscopy (XPS), high-energy grazing-incidence diffuse X-ray scattering (GIXRD) and scanning electron microscopy (SEM). High-photon-energy experiments facilitated the monitoring of changes within the few nm to submicron depth from the surface. A gradual dissolution of $Nb_2O_5$, $NbO_2$ and $NbO$ layers at the surface upon annealing, followed by oxygen diffusion into Nb, allied with the lack of evidence of nitrogen diffusion in infusion and higher-temperature treatment is discussed.

## II. EXPERIMENTAL METHODS

Mechanically polished niobium (100) samples (10 mm diameter, 2 mm height) oriented to better than 0.1° were employed for this study. Unless stated otherwise, the samples were annealed at 2000°C for 6 hours by induction heating in ultra-high vacuum (UHV) at a starting pressure of ~$10^{-10}$ mbar and unavoidably exposed to air for several days before the experiments. This procedure ensures that the Nb single crystals are as clean as possible in the bulk.

A four-circle and a six-circle[17–19] diffractometer with Mo-$K_\alpha$ and Cu-$K_\alpha$ sources, respectively, were employed for *in-situ* X-ray reflectivity measurements of the 800°C annealed and nitrogen-infused samples. The stepwise annealing of the specimens was carried out at the beamline ID31 at the ESRF in Grenoble, France, and followed by *in-situ* high-energy GIXRD and XRR measurements using a photon energy of 70 keV while the data were collected using two separate 2D pixel detectors. The use of high photon energies provides fast access to large areas in reciprocal space, excluding the need of time-consuming sample and detector movements[20,21]. For all aforementioned studies, the samples were mounted in a dedicated UHV setup, allowing direct measurements at different temperatures and gas conditions. For all experiments involving the presence of nitrogen, nitrogen gas (99.999% purity) was inserted in the chamber through a leak valve until the system reached a pressure of $3.3 \cdot 10^{-2}$ mbar after which all the valves were closed and the chamber was kept under static nitrogen atmosphere. XPS measurements were carried out separately using a lab-based system with monochromatic Al$K_\alpha$ radiation and the spectra were recorded in fixed transmission mode with an energy resolution of 0.4 eV[19]. A high resolution field-emission scanning electron microscope (SEM) was employed to characterize the surface morphology of the specimens.

## III. RESULTS AND DISCUSSION

### 800°C ANNEAL IN UHV

During standard cavity preparation, an 800°C annealing in high-vacuum is performed for a few hours to lower the inherent hydrogen content in the material acquired during manufacturing[1,22]. The presence of hydrogen can lead to the formation of niobium hydrides at the cavity surface, especially during the cooldown phase, leading to a poor performance[9]. A comparison between an "as-received" sample and after 800°C annealing for 1 hour at $10^{-7}$ mbar measured by XRR as well as the derived electron density-profiles are shown in FIG. 1a and FIG. 1b, respectively. At room temperature, the natural oxide layer can be described best (from surface to the metal/oxide interface) using a three-layer model composed of $Nb_2O_5$, $NbO_2$ and NbO, with individual thicknesses of 2.90 nm, 0.60 nm and 1.20 nm, respectively (error bar of $\pm$ 0.1 nm)[23]. The total thickness of the natural oxide layer is 4.71 nm while the roughness varies from 0.2 – 0.4 nm for the oxide/metal interfaces while being close to 1.25 nm for the vacuum/$Nb_2O_5$ interface. After annealing to 800°C, both $Nb_2O_5$ and $NbO_2$ layers reduced completely in favor of a 1.03 nm thick NbO layer while the interface roughness between NbO/Nb increased to 0.82 nm.

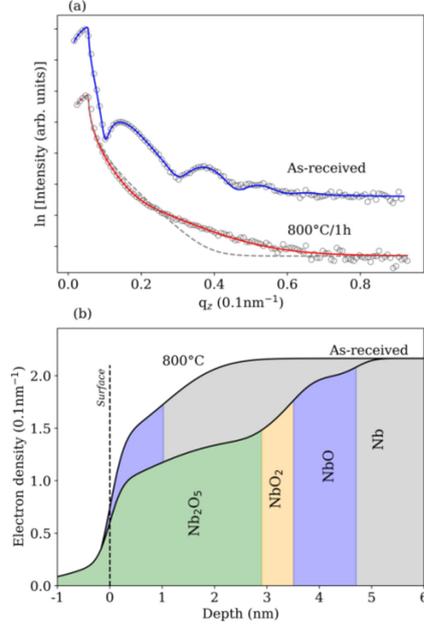

FIG. 1. (a) X-ray reflectivity data measured with Mo $K_\alpha$ radiation (open circles) with respective fits (solid lines) for a Nb(100) single crystal "as-received" and after 800°C annealing in UHV for 1 hour. The dashed line represents the simulated curve for Nb surface without any surface oxide. The scans are offset in the y direction for better visibility. (b) Electron-density profiles for both sets of data obtained from the fits.

It is important to note here that the surface of Nb is not oxide-free after such an annealing, in contrast to what is reported in [2], but a NbO layer remains at the surface. Similar effects were observed for XRR measurements with the Nb(110) surface, where NbO was the only natural oxide remaining after annealing at 300°C in UHV for 50 min, forming an epitaxial (111) oriented NbO layer[23]. The transition from $Nb_2O_5$/$NbO_2$/NbO upon annealing into NbO was also previously confirmed by XPS characterization. Moreover, Auger spectroscopy revealed that clean Nb surfaces could only be achieved after annealing from 1800°C to 2000°C[24], in agreement with the high-resolution STM studies of similarly prepared Nb surfaces[25,26].

For the (100) surface, XPS measurements on single-crystalline Nb etched by buffered chemical polishing (BCP) and annealed at 430K and 540K in UHV, showed predominantly a transition from $Nb_2O_5$ and $NbO_2$ to $Nb_xO$ (0.4 < x < 2) upon annealing up to temperatures close to 265°C , together with an increase in surface roughness[27]. Furthermore, the presence of epitaxial NbO was confirmed by means of grazing-incidence X-ray diffraction (GIXRD) for cavity-grade Nb (100) specimens annealed at 900°C in high-vacuum and treated with nitrogen at the same temperature[10].

Knowing that the release of oxygen by Nb in UHV does not occur below 1600°C[28], it can be concluded that the available oxygen from the partially dissolved natural oxide layer diffuses into the Nb matrix. As will be shown later, for lower annealing temperatures, oxygen diffusion into Nb was confirmed by grazing-incidence X-ray diffuse scattering measurements, similar to what was observed for the (110) surface[23].

## NITROGEN INFUSION

In order to achieve high quality factors ($Q_0 \sim 6 \cdot 10^{10}$) with accelerating gradients up to 45 MVm$^{-1}$, cavities are subjected to the so-called infusion process, which consists of the aforementioned 800°C anneal in high-vacuum followed by a decrease in temperature to 120°C, at which point nitrogen is introduced into the system, maintaining a constant pressure of $3.3 \cdot 10^{-2}$ mbar (25 mTorr) for 48 hours[2]. FIG. 2a and 2b show *in-situ* XRR data measured at each step of the infusion treatment as well as the fitted electron density profiles. At room temperature, before the treatment, the surface displays a similar configuration to the "as-received" sample, with the natural oxides following the order $Nb_2O_5$, $NbO_2$ and $NbO$ from the surface to the metal/oxide interface. Individual thicknesses of 1.27 nm, 0.17 nm and 1.14 nm for $Nb_2O_5$, $NbO_2$ and $NbO$, respectively, were extracted from the fit. All interfaces showed roughness between 0.1 – 0.4 nm. All fits of this sample show a discrepancy in the first few data points due to the surface roundness and faceting caused by the 2000°C anneal performed prior to the experiments[29].

Upon annealing at 800°C for 2 hours, similar to the previous results, the surface was composed of a single NbO layer with thickness of 1.33 nm and slightly superior interface roughness in the range 0.3 – 0.6 nm. While the sample was kept at 120°C, nitrogen was inserted in the chamber until the pressure reached $3.3 \cdot 10^{-2}$ mbar. The chamber valves were then closed, keeping a static nitrogen pressure. The XRR measurements were performed continuously during the total treatment duration (48 hours); however, few differences were observed between the curves. After 1 hour of nitrogen treatment, the XRR curves were essentially identical to the curves obtained after annealing at 800°C for 2 hours. The surface still displayed only a NbO layer with thickness of 1.33 nm. After 48 hours in nitrogen atmosphere the thickness of the NbO layer decreased to 1.20 nm with similar interfacial roughness as the previous step, between 0.3 – 0.6 nm.

Interestingly, no evidence for either a nitrogen-rich or a nitride layer was observed with prolonged exposure of the Nb specimen to a nitrogen atmosphere at 120°C, but rather only the progressive dissolution of the remaining NbO layer at the surface. It is known that a passivating layer of NbO can be achieved after oxidation of Nb thin films at 340°C in an oxygen atmosphere[30], which could in this case hinder nitrogen diffusion. Such a passivating layer was not present in the sample under test, as confirmed by the regrowth of the oxides $NbO_2$ and $Nb_2O_5$ after it was re-exposed to air. Moreover, in contrast to what was observed for cavity-grade Nb upon high-temperature annealing, where rectangular and triangular-shaped precipitates were observed at the surface[10,14], no precipitates were detected by SEM after infusion (see FIG. S1a). This indicates that the initial material purity may also play a role in the formation of such features on the surface.

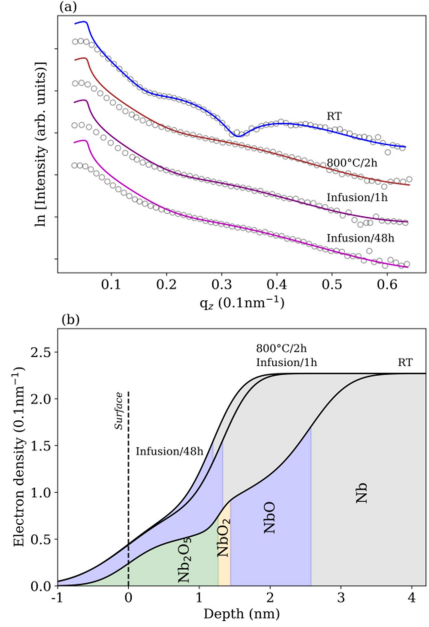

FIG. 2. (a) X-ray reflectivity data measured with CuK$_\alpha$ radiation (open circles) with respective fits (solid lines) for various steps of the infusion procedure. The sample curvature compromises the fit in the first few data points. The scans are offset in the y direction for better visibility. (b) Electron density profiles for each step. The extracted densities are lower than the expected bulk values for the niobium oxides and metallic niobium due to the aforementioned sample curvature.

Although SIMS measurements are difficult when dealing with nitrogen, as well as suffering from an inability to define the exact location of the oxide/metal interface[31], nitrogen incorporation in Nb after after exposure to 3.3·10$^{-2}$ mbar nitrogen at 120 °C[2,16] has been reported. The reported concentrations should have been easily detectable by XRR measurements. One hypothesis to explain such discrepancies is the absence of grain-boundaries in the samples studied in this work, which can act as enhanced diffusion sites for nitrogen atoms[32]. Results from Nb 3d XPS spectra for polycrystalline Nb subjected to annealing cycles of 800°C for 3 hours in high-vacuum followed by 48 hours at 140°C with a nitrogen pressure of 3.3·10$^{-2}$ mbar, were interpreted as displaying NbN$_{(1-x)}$O$_x$ at a peak position of 203.2 eV[16]. Such a peak was also attributed by other authors[33,34] as belonging to NbC as well as NbO, therefore an analysis of the C 1s, N 1s and O 1s edges is necessary to understand the system fully. This is presented below.

## STEPWISE ANNEALING AND INTERSTITIAL X-RAY DIFFUSE SCATTERING

Interstitial atoms, such as oxygen and nitrogen, can occupy octahedral sites in the Nb bcc lattice[35,36]. Together with the natural mechanical instability along the [111] direction of the lattice leading to the ω phase[35,37], oxygen and nitrogen-induced local distortions can produce diffuse scattering maxima at $\boldsymbol{q}^* = \frac{2}{3}(1\ 1\ 1)$ and space group Im-3m symmetry equivalent places in reciprocal space[38]. This diffracted intensity, in a regime of random defect concentration, can be written as[39]:

$$I_D \propto c(1-c)S(\boldsymbol{q}) \qquad \text{(Eq. 1)}$$

Where $c$ is the concentration of the defect which is responsible for the intensity while $S(\mathbf{q})$ is the square of the scattering amplitude of a single defect.

For the experiment, the sample was placed inside a dedicated UHV chamber and mounted on the diffractometer in grazing-incidence geometry. By tuning the incident and exit angles of the X-rays and therefore the so-called scattering depth[40,41] $\Lambda$, which determines the depth from whence the observed intensity originates. The probed depth can be adjusted down from few nm up to the µm scale[42].

Table I. Temperature, total time and fitted decay lengths obtained by grazing incidence diffuse scattering measurements.

| Temperature (°C) | Total time (hh:min) | Decay length (nm) |
|---|---|---|
| RT | -- | 0.969 |
| 120 | 13:45 | 0.989 |
| 200 | 14:00 | 1.970 |
| 250 | 15:20 | 2.713 |
| 250 + $N_2$ | 14:20 | 3.576 |

Contrary to the (110) surface, where the [111] direction resides in-plane, for the (100) surface the diffuse maxima has an out-of-plane component. This makes the tuning of the X-ray exit angle impossible during the experiment. However, the probed depth can still be changed by modifying the X-ray incidence angle. Taking into account the sample curvature, it is possible to calculate an effective scattering depth ($\Lambda_{eff}$) for each measured step[42].

The evolution of the diffuse intensity was followed *in-situ* during stepwise annealing in UHV ($10^{-8}$ mbar) as well as in a nitrogen atmosphere. Employing high-energy grazing incidence X-ray diffraction in combination with a large 2D detector made it possible to sample simultaneously large areas of reciprocal space[20,21]. Starting from room temperature (RT), the sample was annealed at 120°C, 200°C and 250°C, sequentially. At 250°C, nitrogen was inserted into the chamber until the pressure reached $3.3 \cdot 10^{-2}$ mbar. Similar to the previous experiments, the $N_2$ atmosphere was kept under static conditions. Furthermore, the sample temperature was increased to 500°C in the $N_2$ atmosphere as a last annealing step before air exposure. Together with diffuse scattering measurements, XRR curves were obtained for each temperature, providing information about the surface oxides. Table I summarizes the annealing temperatures and times at each step.

The thickness and roughness of the oxide layers during each step of annealing are presented in FIG. 3, while FIG. 4 shows the related *in-situ* XRR curves and respective electron-density profiles extracted. The natural oxide layer was composed, beginning at the surface and ending at the metal/oxide interface, of $Nb_2O_5$, $NbO_2$ and $NbO$. The outmost oxide, $Nb_2O_5$, gradually decreases in thickness upon annealing up to 200°C and vanishes at 250°C. Until 200°C, $NbO_2$ followed the same trend as $Nb_2O_5$. However, upon further annealing it experiences a slight increase in thickness of about 0.09 nm, remaining almost constant thereafter. Hardly affected by annealing at 120°C, $NbO$ grows at the expense of the other oxides after annealing at 200°C and starts to dissolve upon further annealing. The substrate roughness is barely affected by annealing at 120°C and shows a steady growth as the temperature is increased. The stepwise dissolution of the natural oxides was also observed for the Nb(110) surface by both XPS and XRR measurements. In contrast to the results shown here, where $NbO_2$ is the most affected oxide during 120°C annealing, having a decrease in thickness of ~0.22 nm, for the (110) surface, $Nb_2O_5$ is the most affected oxide – having a similar decrease of ~0.23 nm after annealing for 5 hours in UHV at 145°C[23].

The behavior of the natural oxides proposed here is similar to what was observed in STM studies of the (100) surface, for which a reduction from $Nb_2O_5$ to $NbO_2$ was observed in form of a ladder-like structure caused by oxygen segregation at the surface during annealing at 580°C in UHV with sub-sequential re-growth of $Nb_2O_5$ after air exposure[43].

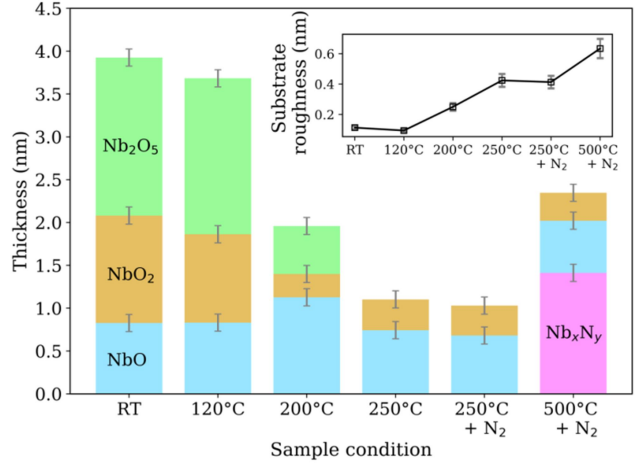

FIG. 3. Thickness and substrate roughness (inset) obtained by XRR in each step of the sample treatment.

After $N_2$ insertion in the system, no new nitride or nitrogen-rich layer was observed at 250°C, but rather only a further decrease in oxide thickness. However, at 500°C a new layer with thickness of 1.41 nm was formed between NbO and Nb. The electron density of this layer is similar to the theoretically calculated value for the β-$Nb_2N$ phase, which is known to precipitate in Nb upon high–temperature treatments in a $N_2$ atmosphere[10,14]. Therefore, it can be concluded that the layer relates to the presence of a possible $Nb_xN_y$ compound underneath the oxide layer, as the presence of typical β-$Nb_2N$ precipitates at the surface was not observed by SEM (FIG. S1b).

The experimental setup for the interstitial induced diffuse-scattering measurements is illustrated in FIG. 5. The sample was rocked around the Nb $hhl$ plane within a range of $\pm 10°$, with a region of interest ($\Delta q_{xy} = \Delta q_z = 0.42$ Å$^{-1}$) set around the theoretically calculated diffuse maxima at $\boldsymbol{q}^* = \frac{4}{3}(1\ 1\ 1)$, depicted by a blue circle in the reciprocal-lattice sketch. The region of interest was integrated in rotation for each incident angle and plotted against the corresponding scattering depth $\Lambda_{eff}$. The resulting curve was fitted using an exponential decay model for the interstitial concentration such as:

$$c(z) = ae^{-z/\tau} + d \qquad (Eq.\ 2)$$

where $a$ (pre-exponential factor), $\tau$ (decay length, defined as where $c(z) = 1/e$) and $d$ (*bulk concentration*) are the fitted parameters. $z$ is the depth, with zero being the vacuum/oxide interface. Thermal diffuse-scattering contributions as well as optical factors[41,42] were also taken in account by the fitting routine. A set of integrated detector images obtained at different temperatures are displayed in FIG. 6 for a fixed incident angle of 0.04°, corresponding to a scattering depth $\Lambda_{eff}$ of ~100 nm.

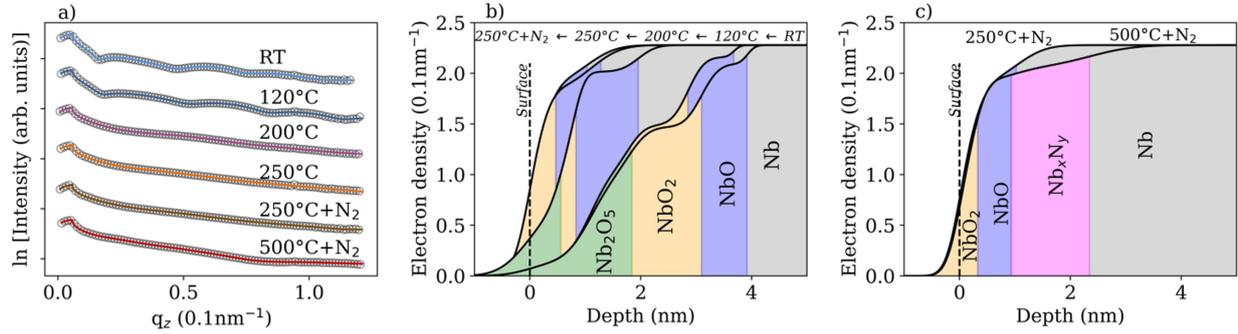

FIG. 4. (a) X-ray reflectivity data (open circles) measured with photon energies of 70 keV with the respective fits (solid lines). The scans are offset in y-direction. (b) Electron-density profiles obtained from the fits for the data collected from RT up to the addition of $N_2$ at $3.33 \times 10^{-2}$ mbar at 250°C. (c) Comparison between the electron density profiles obtained at 250°C and 500°C in $N_2$ atmosphere.

FIG. 7(a) shows the diffuse scattering data collected from RT up to 250°C in a $N_2$ atmosphere. The interstitial concentration profile extracted from the fits is presented in FIG. 7(b). An increase in the decay length of subsurface oxygen, associated with a decrease of the initial concentration (at z = 0) was observed as the temperature was increased, as seen in the inset of FIG. 7(b). This behaviour can be understood as the progressive diffusion of oxygen liberated from the oxide layers, which in turn decrease in total thickness. Furthermore, as the temperature increases, the diffusion length of the liberated oxygen in Nb increases drastically. The interstitial concentration in the first 10 nm is the highest during annealing at 120°C, which agrees with the 12 nm of calculated diffusion length of oxygen in pure Nb[28]. Compared to 120°C, annealing at 200°C liberates more oxygen in the Nb matrix and enables it to diffuse further. This is seen by the increase in the bulk concentration from $\sim 3 \cdot 10^{-9}$ to $\sim 3 \cdot 10^{-8}$ as well as the decay-length increase from 0.98 nm to 1.97 nm. On the other hand, the subsurface concentration in the first tens of nm decreases. The same trend continued during annealing at 250°C, increasing even more the decay length up to 2.71 nm and the bulk concentration to $\sim 5 \cdot 10^{-6}$. With $N_2$ insertion in the system one would expect the amount of interstitials in the first 10 nm to increase, since the diffusion length of N in pure Nb at 250°C for 12 hours is about 11.2 nm[28]. However, this is not the case. The interstitial concentration within the first few nm decreased, while the decay length increased to 3.57 nm. The bulk concentration remained essentially the same. This means that annealing in a $N_2$ atmosphere at 250°C promoted only further diffusion of oxygen originating from the surface oxides into Nb with no evidence of nitrogen incorporation in the metallic Nb matrix. The corresponding decay lengths for different annealing steps are summarized in Table I.

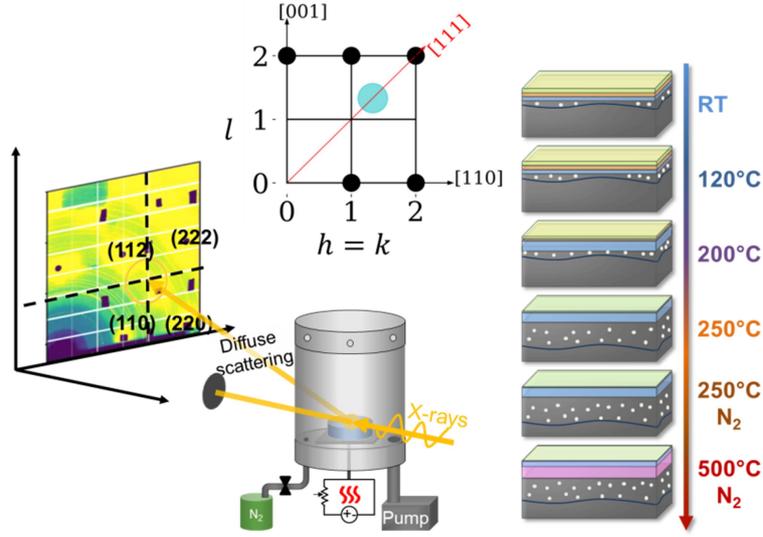

FIG. 5. The high-energy beam falls on the sample surface at grazing incident angles and the diffraction patterns are obtained by a 2D detector while the sample undergoes different annealing treatments under UHV and $N_2$. The natural oxide layers on the sample gradually dissolve and the liberated oxygen atoms diffuse into the Nb matrix. In the reciprocal lattice sketch of the ***hhl*** Nb plane, the characteristic positions for the interstitial diffuse scattering in Nb along the [111] direction are depicted in blue, while Nb reflections are shown in black. At 500°C under a $N_2$ atmosphere, a $Nb_xN_y$ layer is formed underneath the remaining niobium oxides.

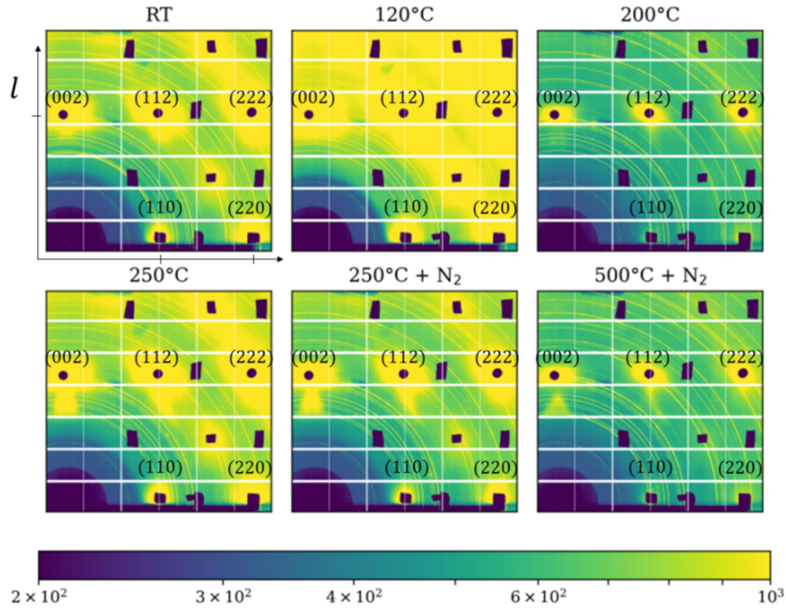

FIG. 6. Integrated images collected for different temperatures with an incident angle of 0.04°, corresponding to a scattering depth $\Lambda_{eff}$ of ~100 nm. At 120°C, a clear increase in diffuse intensity is observed due to the presence of subsurface oxygen. Lead (Pb) pieces block the Nb Bragg peaks. Powder rings originating from the beryllium (Be) windows of the chamber are also observed.

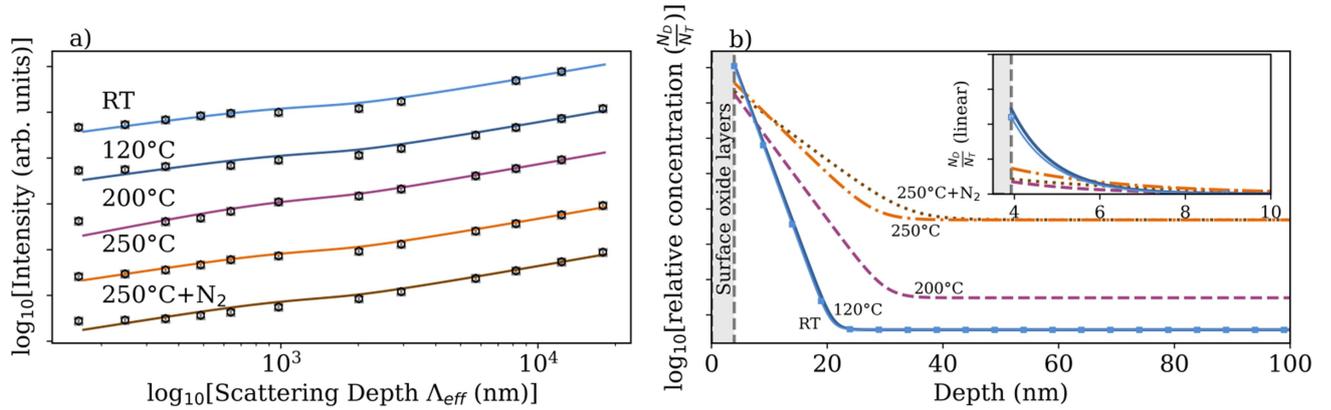

FIG. 7. (a) Diffuse scattering intensities obtained for each temperature step (squares) and the respective fit (solid line). The curves are offset in y-direction. (b) Interstitial concentration obtained from the fit plotted in log scale and the concentration in linear scale (inset). The profiles are represented by connected squares (room temperature), solid (120°C), dashed (200°C), dot-dashed (250°C) and dotted lines (250°C under $N_2$ atmosphere).

High-resolution XPS measurements were performed after the sample was re-exposed to air for ten days. Nb 3d spectra observed for the normal, as well as grazing-exit mode, are shown in FIG. 8a and b, which confirm the regrowth of the natural oxide layers previously dissolved due to the stepwise UHV annealing. The characteristic doublets of metallic Nb are observed at the binding energies of 202.0 eV and 204.7 eV. The top surface oxide layer on niobium is consisting of $Nb_xO$, $NbO$, $NbO_2$ and $Nb_2O_5$, for which the binding energies exceed to metallic niobium by 0.7 eV, 1.9 eV, 3.0 eV and 5.0 eV, respectively. A strong contribution from NbC is suggested by the presence of $3d_{5/2}$ and $3d_{3/2}$ peaks at 203.3 eV and 206.0 eV, respectively, and confirmed by the presence of a peak in C 1s spectra at 282.6 eV assigned to niobium carbon bonding[44] (FIG. S2). Furthermore, a contribution from two peaks at binding energies of 204.3 eV and 207.0 eV is detected, which can be attributed to a $NbN_xO_y$ phase, supported by the presence of a peak in N 1s line spectra (FIG. 8c and d) at the binding energy of ~397.5 eV, which corresponds to niobium bound to nitrogen in niobium nitrides and oxynitrides as reported for the nitrogen-doped niobium samples[10]. The higher binding energy peak in N 1s spectra at ~ 400 eV is due to the adsorbed nitrogen on the surface. The intensity of the lower binding energy peak in N 1s spectrum is significantly suppressed when measured at the grazing-exit angle given in FIG. 8d, where the majority of detected photo electrons correspond to the top surface layer. Moreover, the peaks corresponding to the $NbN_xO_y$ phase in normal-exit Nb 3d spectrum in FIG. 8a are not observable in the grazing-exit Nb 3d spectrum in FIG. 8b. These results indicate that the nitrogen-containing phase does not reside on the top layer but close to the oxide-metal interface, which is in-line with the *in-situ* XRR measurements on the same sample at 500°C in nitrogen atmosphere.

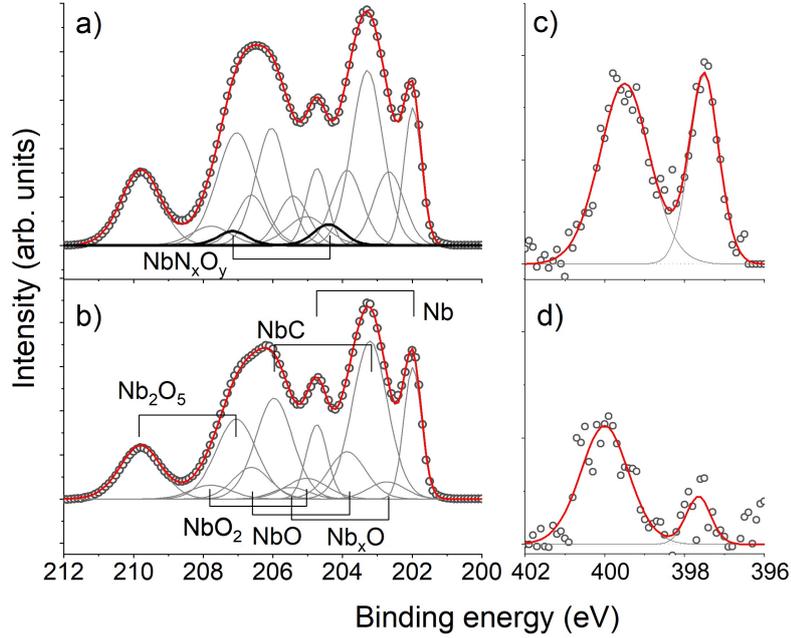

FIG. 8. Deconvolution of Nb 3d and N 1s lines in normal exit mode (a,c) and with 70° exit angle (b,d). Phases corresponding to niobium, niobium oxides and niobium carbides are represented in gray. Contributions from niobium nitrides are shown in black.

## IV. CONCLUSIONS

The effect of different thermal treatments that have been recently applied to Nb SRF cavities was studied in the near-surface region of single-crystalline Nb(100). *In-situ* XRR revealed that annealing to 800°C in UHV reduces the natural oxide layer composed of $Nb_2O_5$, $NbO_2$ and NbO into solely NbO; the natural oxide layer is regrown after exposure to air. After nitrogen exposure at 120°C and $3.3·10^{-2}$ mbar, neither a nitrogen-rich layer nor interstitial nitrogen were detected. *In-situ* high-energy X-ray diffuse scattering combined with XRR measurements during stepwise annealing in UHV up to 250°C demonstrated that the partial dissolution of the natural oxides leads to diffusion of oxygen into Nb. At 120°C, interstitial oxygen is mostly present within the first 10 nm, while temperature increase leads to further diffusion into the *bulk*. When the sample was exposed to $3.3·10^{-2}$ mbar of $N_2$ at 250°C, no evidence of nitrogen presence either in the form of a nitrogen-containing layer or as interstitial nitrogen was detected. Annealing at 500°C resulted in the formation of a $Nb_xN_y$ layer underneath the remaining $NbO_2$ and NbO at the surface. *Ex-situ* photoelectron measurements confirm the presence of Nb-N bonding as well as the regrowth of the natural oxide layer at the surface upon exposure to air. These results suggest that nitrogen may not be responsible for the increased performance of SRF cavities after low-temperature annealing, as higher temperature treatments are necessary for nitrogen to react effectively with Nb. A rearrangement of the oxygen interstitial profile takes place during the low-temperature annealing of Nb in a nitrogen atmosphere, which is suggested in the literature to be potentially partially responsible for cavity improvements[2]. Further studies concerning the effect of grain-boundaries as well as other crystal orientations, are recommended in order to elucidate the apparent improvement in the performance of SRF cavities subjected to low-temperature thermal treatments.

## ACKNOWLEDGEMENTS

Authors declare that no conflict of interest exists. SEM analysis by S. Kulkarni and A. Jeromin is acknowledged. The authors thank T. Meisner and A Weible for the UHV-induction annealing of the Nb crystals at MPI for intelligent systems. Authors GDS and BF acknowledge the funding from the BMBF grant no. 05H15GURBB. Author GDS would like to thank K. P. Furlan for the revision of the presented figures. All figures and pictures by the author(s) under a CC BY 4.0 license.

174 (2008).

[34] A. Daccà, G. Gemme, L. Mattera, and R. Parodi, Appl. Surf. Sci. **126**, 219 (1998).

[35] H. Dosch, A. v. Schwerin, and J. Peisl, Phys. Rev. B **34**, 1654 (1986).

[36] M.S. Blanter and A.G. Khachaturyan, Metall. Trans. A **9**, 753 (1978).

[37] R.P. Kurta, V.N. Bugaev, A. Stierle, and H. Dosch, J. Phys. Condens. Matter **20**, 275206 (2008).

[38] H. Dosch and J. Peisl, Phys. Rev. B **32**, 623 (1985).

[39] M.A. Krivoglaz, *X-Ray and Neutron Diffraction in Nonideal Crystals* (Springer Berlin Heidelberg, Berlin, Heidelberg, 1996).

[40] H. Dosch, Phys. Rev. B **35**, 2137 (1987).

[41] H. Dosch, B.W. Batterman, and D.C. Wack, Phys. Rev. Lett. **56**, 1144 (1986).

[42] H. Dosch, *Critical Phenomena at Surfaces and Interfaces : Evanescent X-Ray and Neutron Scattering* (Springer-Verlag, 1992).

[43] Y. Li, B. An, S. Fukuyama, K. Yokogawa, and M. Yoshimura, Mater. Charact. **48**, 163 (2002).

[44] A. Darlinski and J. Halbritter, Surf. Interface Anal. **10**, 223 (1987).
15